\documentclass[twocolumn,showpacs,amsmath,amssymb]{revtex4}
\usepackage{graphicx}
\begin{document}

\title{Probing the superconducting pairing symmetry from spin excitations in BiS$_2$ based superconductors }

\author
{Tao Zhou$^{1,2}$ and Z. D. Wang$^{2}$}

\affiliation{$^{1}$College of Science, Nanjing University of Aeronautics and Astronautics, Nanjing 210016, China\\
$^{2}$Department of Physics and Center of Theoretical and Computational Physics, The University of Hong Kong,
Pokfulam Road, Hong Kong, China}

\date{\today}
\begin{abstract}
Starting from a two-orbital model and based on the random phase approximation, spin excitations in the superconducting state of the newly discovered BiS$_2$ superconductors with
three possible pairing symmetries are studied theoretically.
We show that spin response is uniquely determined by the pairing symmetry.
Possible spin resonance excitations might occur for the $d$-wave symmetry at an incommensurate momentum about $(0.7\pi,0.7\pi)$. For the $p$-wave symmetry
 the transverse spin excitation near $(0,0)$ is enhanced. For the $s$-wave pairing symmetry there is no spin resonance signature.
These distinct features may be used for probing or determining the
pairing symmetry in this newly discovered compound.
\end{abstract}
 \maketitle

Very recently, new BiS$_2$ based superconducting (SC) materials were discovered~\cite{yos,yosh,sato}. The materials reported so far include Bi$_4$O$_4$S$_3$~\cite{yos}, LaO$_{1-x}$F$_x$BiS$_2$~\cite{yosh}, and NdO$_{1-x}$F$_x$BiS$_2$~\cite{sato}. These materials share some common features with cuprate-based and iron-based high-Tc materials, e.g., they are layered materials with BiS$_2$ layers being the SC plane. Superconductivity is realized by adding defects into the parent systems. Thus this family might open a new door in the studies of unconventional superconductivity, providing a new insight into the origin of high-Tc superconductivity.
 Actually, it has already attracted considerable interest in investigating the physical properties of them~\cite{usui,sli,sing,awa,kot}.

 Understanding the pairing mechanism and probing the pairing symmetry are two of most important issues in the studies of SC materials. Now it is believed that the spin fluctuation should be responsible for the SC pairing in many families of unconventional SC materials, including heavy fermion
materials, cuprates and iron-based ones~\cite{scal}. For BiS$_2$ based materials, in was found from band calculation that there exists a good nesting of Fermi surface, thus the spin fluctuation should also be considerably strong~\cite{usui}. On the other hand, it was proposed in Ref.~\cite{sli} that the SC pairing is strong and exceeds the limit of the phonon mediated picture. Therefore, the spin fluctuation is a good candidate to be responsible for the SC pairing in this family. In the mean time, the results of the spin susceptibility are usually phase sensitive due to the existence of coherence factor, thus
it has been a powerful tool for probing the pairing symmetries~\cite{jxli,jxli1,mai,seo}. At last, many exotic features, e.g., the spin resonance and possible incommensurate (IC) spin excitation, have attracted broad interest in previous studies of various families of unconventional superconductors~\cite{morr,zhou,gao}.
 Therefore, at this stage it is highly desirable and of importance to study the spin excitations in the BiS$_2$-based materials and to look into the pairing symmetry theoretically.

Band structure for LaO$_{1-x}$F$_x$BiS$_2$ sample was obtained from first principles calculation in Ref.~\cite{usui}. A simplified two-orbital (considering Bi 6p orbital) model was proposed by neglecting the interlayer coupling and the bands far away from Fermi surface~\cite{usui}. In this paper, by employing this two-orbital model, we study theoretically the spin
fluctuations based on the random phase approximation (RPA).
Three possible SC pairing symmetries ($d$-wave, extended $s$-wave, and $p$-wave) have been considered to study the spin response in the SC state.
We find that the spin response is enhanced at an IC momentum near $(\pi,\pi)$ [or $(0,0)$] for the $d$-wave [or $p$-wave] pairing symmetry, while there is no significant feature of the spin excitations for the $s$-wave pairing.
These could be
used to test the pairing symmetry experimentally.

Our starting model is expressed as,
\begin{equation}
H=\sum_{{\bf ij\sigma},\mu\nu} t_{{\bf ij},\mu\nu}c^{\dagger}_{{\bf i}\mu\sigma}c_{{\bf j}\nu\sigma}+H_\Delta+\sum_{{\bf i}\mu} U n_{{\bf i\uparrow}\mu}n_{{\bf i\downarrow}\mu},
\end{equation}
where $\mu,\nu$ are the orbital indices. In the present work, we use the two orbital model with the hopping constants being from Ref.~\cite{usui}.
$H_\Delta$ is the SC pairing term.
 The last term is the on-site electron interaction and will take effect in the RPA framework. Usually the interaction term should include the onsite repulsive interaction, and the Hund coupling term. Both terms play a similar role and enhance the spin excitation after the RPA correction. Thus we here consider only the onsite intraorbital interaction term. We have also checked numerically that the main results do not change when an additional inter-orbital interaction is added.

We define the single-particle normal Green's function $G$ and anomalous Green's function $F(\widetilde{F})$ with $G^{\sigma}({\bf k},i\tau)_{\mu\nu}=-\langle T_{\tau} c_{\mu\sigma}({\bf k},-i\tau)c^{\dagger}_{\nu\sigma}({\bf k},0)\rangle$, $F^{\sigma}({\bf k},i\tau)_{\mu\nu}=-\langle T_{\tau} c_{\mu\sigma}({\bf k},-i\tau)c_{\nu,-\sigma}({-\bf k},0)\rangle$,
and $\widetilde{F}^{\sigma}({\bf k},i\tau)_{\mu\nu}=-\langle T_{\tau} c^{\dagger}_{\mu\sigma}({\bf k},-i\tau)c^{\dagger}_{\nu,-\sigma}({-\bf k},0)\rangle$.

Then the bare spin susceptibility can be defined as,
\begin{eqnarray}
{\chi}^{zz,+-}_{0,ij}(q)=&-\frac{1}{2\beta N}\sum_{k\sigma}[{G}^{\sigma}_{ij}(k){G}^{\sigma^{\prime}}_{ij}(k+q)\nonumber\\&- \sigma\sigma^{\prime}\widetilde{F}^{\sigma}_{ij}(k){F}^{\sigma^{\prime}}_{ij}(k+q)],
\end{eqnarray}
Here $k=({\bf k},i\omega_n)$ and $q=({\bf q},i\omega)$.
$A(k)$ ($A=G,F,\widetilde{F}$) are the Fourier transformations for $A({\bf k},{i\tau})$.
$\sigma^{\prime}=\sigma$ for $\chi^{zz}_0$ and $\sigma^{\prime}=-\sigma$ for $\chi^{+-}_0$, respectively.

The anomalous Green's function in Eq.(2) satisfy $\widetilde{F}^{\sigma}_{ij}(k)={F}^{-\sigma}_{ji}(-k)=-{F}^{\sigma}_{ji}(k)$.
For the spin singlet pairing, ${F}^{\sigma}_{ij}(k)=-{F}^{-\sigma}_{ij}(k)$,
thus the formulas for $\chi^{zz}$ and $\chi^{+-}$ are the same.
For the triplet pairing, we have ${F}^{\sigma}_{ij}(k)={F}^{-\sigma}_{ij}(k)$,
so the spin excitations for $\chi^{zz}$ and $\chi^{+-}$ are different with an additional minus sign in front of the anomalous part of spin susceptibility.

The spin susceptibility within RPA is given by
\begin{equation}
\hat{\chi}({\bf q},\omega)=\frac{\hat{\chi}_0({\bf q},\omega)}{{\bf I}-U{\bf I}\hat{\chi}_0({\bf q},\omega)},
\end{equation}
where ${\bf I}$ is a unit matrix. The total (physical) spin susceptibility $\chi$ is $\sum_{ij}\chi_{ij}$.

\begin{figure}
\centering
  \includegraphics[width=6.3cm]{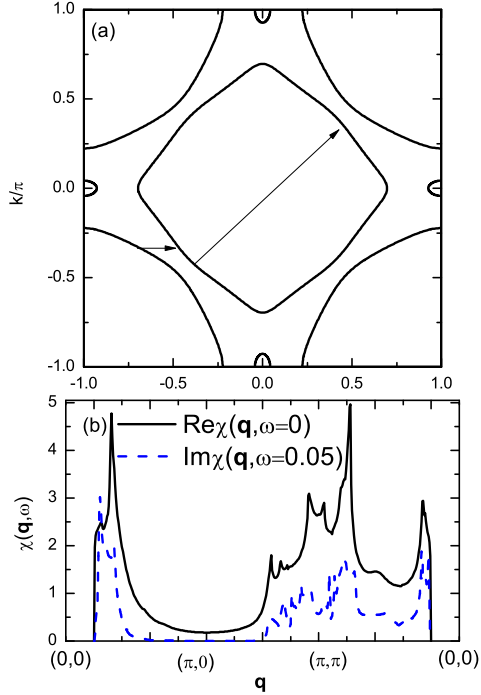}
\caption{(Color online) (a) The normal state Fermi surface. (b) The normal state spin excitations along different momentum cut.}
\end{figure}

Before we present our numerical results, we need to determine the doping density. For the Bi$_4$O$_4$S$_3$ sample, the band calculations suggest that the electron doping is $x=0.5$~\cite{yos}. For LaO$_{1-x}$F$_x$BiS$_2$ sample, superconductivity is observed for $x\geq0.5$ and it was proposed that the optimal doping density is slightly above 0.5~\cite{yosh}. For the NdO$_{1-x}$F$_x$BiS$_2$ sample, the range of SC state seems to be broad with superconductivity being observed for $0.1<x<0.7$~\cite{sato}. We also noted that the Fermi surface topology changes significantly as the doping increases from $0.25$ to $0.5$~\cite{usui}, thus the pairing symmetry and the spin excitations might be quite different for different doping densities  for the NdO$_{1-x}$F$_x$BiS$_2$ sample, while this issue was not concerned in the present work. We focus our studies on the optimal doped sample with $x=0.55$ used throughout the present work. Other parameters are chosen as $U=1.5$, $\Delta_0=0.06$.

The normal state Fermi surface and spin susceptibility as a function of momentum $\bf q$ are plotted in Fig.1(a) and
 Fig.1(b). As is seen, the spin excitations are mainly concentrated near the momentum $(0,0)$ and $(\pi,\pi)$. We also find that the low energy spin excitations are incommensurate, with the maximum spin excitations being at $(0,\pm \delta)$ and $(\pi\pm\delta,\pi\pm\delta)$. The origin of these features is explained well based on the Fermi surface nesting picture. As seen in Fig.1(a),
 since the band structure includes two quasi-one dimensional band, the Fermi surface exhibits a good nesting feature with the vectors shown in Fig.1(a) by arrows~\cite{usui}.

 The pairing symmetry can be estimated by assuming that SC pairing is mediated by the spin fluctuation~\cite{tak,kur,zjyao}. Generally, if the pairing potential around the momentum $\bf q$ determine the SC pairing, the relation $\Delta({\bf k})=-\Delta({\bf k}+{\bf q})$ should be satisfied with $\bf k$ being the wave vector on the Fermi surface. With a weak-coupling analysis~\cite{kur}, we can obtain three possible pairing symmetries, displayed in Fig.2. For a singlet pairing symmetry, their exist two possible pairing symmetries, as shown in Fig.2(a) and 2(b). The $d$-wave symmetry with $\Delta({\bf k})=\Delta_0(\cos k_x-\cos k_y)/2$ shown in Fig.2(a) is caused by the $(\pi,\pi)$-spin excitation.
Meanwhile, both $(0,\pm \delta)$ and $(\pi,\pi)$-spin excitation would give rise to the extended s-wave pairing symmetry with $\Delta({\bf k})=\Delta_0(\cos k_x+\cos k_y)/2$. For the triplet pairing, the $p$-wave pairing with $\Delta_{\bf k}=\Delta_0 \sin (k_x\mp k_y)$ (with $'-,+'$ for orbital 1,2, respectively) is favored.

\begin{figure}
\centering
  \includegraphics[width=8cm]{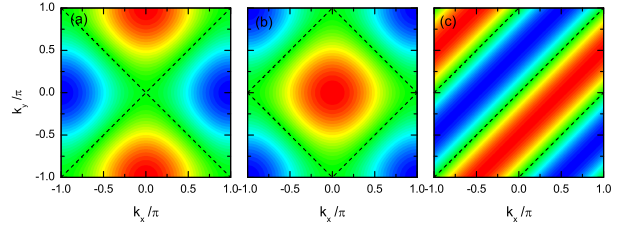}
\caption{(Color online) Three possible SC pairing. (a) $\Delta_{\bf k}\propto \cos k_x-\cos k_y$. (b) $\Delta_{\bf k}\propto \cos k_x+\cos k_y$. (c) $\Delta_{\bf k}\propto \sin (k_x\mp k_y)$.}
\end{figure}

\begin{figure}
\centering
  \includegraphics[width=6.4cm]{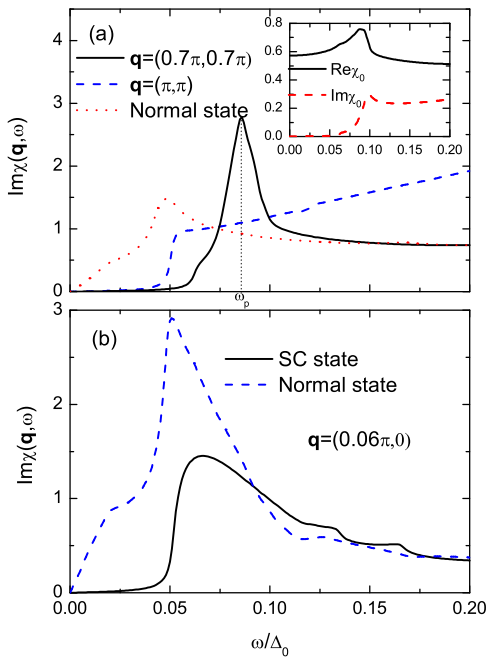}
\caption{(Color online) The imaginary parts of the spin susceptibility as a function of frequency at different momenta for the $d$-wave SC state.
The normal state spin excitations are also displayed for comparison.
The inset of panel (a) shows the bare spin susceptibility.}
\end{figure}

The numerical results of the spin excitations for the $d$-wave pairing symmetry are presented in Fig.3. At the $(\pi,\pi)$ momentum the spin susceptibility increases as the frequency increases and no peak structure is seen. As the momentum is transferred to the IC one, i.e.,
${\bf q}=(\pi\pm\delta,\pi\pm\delta)$ ($\delta$ is the incommensurability),
the peak appears at the frequency $\omega=\omega_p$. Our numerical results also show that $\omega_p$ increases as the incommensurability $\delta$ increases (not shown here). In the meantime, the intensities of the IC peaks also increase as $\delta$ increases. At the momentum about $(0.7\pi,0.7\pi)$, the IC peaks reach the maximum intensity at the frequency about $\omega_p=0.08$. At higher frequencies, spin excitations
at the IC momentum decreases as the frequency increases. As $\omega>0.1$, the commensurate $(\pi,\pi)$ spin excitation is dominant over IC ones, as seen in Fig.3(a). In the meantime, we also looked into the spin excitations around the parallel direction with ${\bf q}=(\pi,\pi\pm\delta)$, and found that the parallel spin excitations are always smaller than the diagonal ones, therefore they are not presented here.

We also plotted the RPA normal state spin susceptibility at the momentum $(0.7\pi,0.7\pi)$ in Fig.3(a). Obviously, the spin excitation is enhanced significantly in the SC state, such that it is quite like a resonance behavior.
Usually the spin resonance occurs for the pole condition of PRA factor and negligibly small bare spin susceptibility.
To check whether it is the resonance or quasi-resonance behavior, we plotted the bare spin susceptibility in the inset of Fig.3(a).
As is seen, at very low energies the imaginary part of bare spin susceptibility is zero due to the presence of the spin gap.
 At the gap edge, there is a
steplike rise of Im$\chi_0$. Due to the
Kramers-Kroenig relations the Re$\chi_0$ develops a peak structure. As a result,
the spin resonance would be expected when $1-U$Re$\chi_0$ equals to zero (or small, corresponding to a small onsite interaction U).
Actually for the present case we can conclude that it is indeed a resonance behavior while $U$ is not very large here so that the resonance peak is not very strong.
It will increase greatly if a larger $U$ is taken into account, but it is not presented in the present work.

\begin{figure}
\centering
  \includegraphics[width=6.3cm]{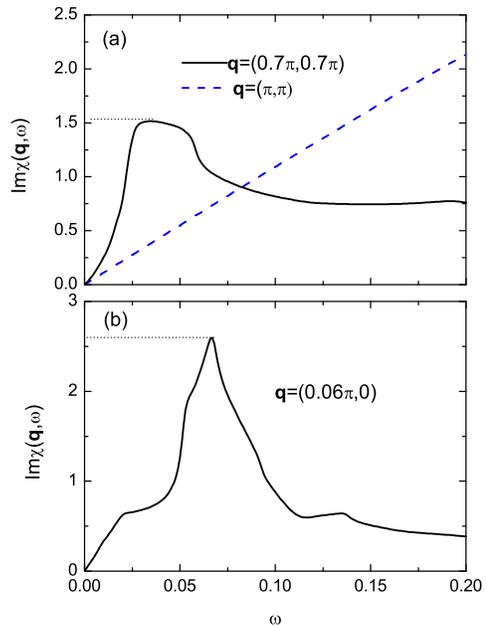}
\caption{(Color online) The imaginary parts of the spin susceptibility as a function of frequency at different momenta for the $s$-wave SC state. }
\end{figure}

For small momenta, the spin excitations are incommensurate for all of the energies $(\omega<0.3)$ we considered, with the IC peaks along the $k_y=0$ direction. We plotted the frequency dependence of RPA spin susceptibility Im$\chi({\bf q},\omega)$ for the momentum ${\bf q}=(0.06\pi,0)$ in Fig.3(b). As is seen, the intensity of the IC peak is suppressed significantly by the SC order. This is due to the presence of the coherence factor in the bare spin susceptibility~\cite{jxli1,mai,seo}. In principle, for spin-singlet pairing symmetry, the spin response is enhanced if $\Delta_{\bf k}\Delta_{{\bf k}+{\bf q}}<0$ is satisfied, otherwise it would be suppressed in the SC state. Thus it is quite reasonable that the spin response is enhanced near $(\pi,\pi)$ and suppressed around small momenta for the $d$-wave pairing symmetry.

The spin excitations for the extended $s$-wave are presented in Fig.4. Fig.4(a) shows the spin excitation near $(\pi,\pi)$. As is shown, the spin susceptibility increases almost linearly as the frequencies at the commensurate momentum. For the IC momentum $(0.7\pi,0.7\pi)$, the spin susceptibility exhibits a broad peak. However,  at this momentum the peak intensity nearly equals to that of the normal state shown in Fig.3(a). Thus there is no spin resonance excitations. We also examine other momenta along various directions and no obvious resonant spin excitation was found. On the other hand, the IC behavior is similar to the case of $d$-wave symmetry. The spin response is incommensurate for low frequencies. The incommensurability decreases as the frequency increases and the spin excitations shift to the commensurate one at the frequency about 0.15.

The spin excitation around a small momentum is displayed in Fig.4(b). The spin excitations do not show much difference compared to the normal state one. Actually, since the nodal lines are close to the Fermi surface, the pairing gap at the fermi surface is quite small. As a result, there is nearly no significant features for the SC spin response.

\begin{figure}
\centering
  \includegraphics[width=6.3cm]{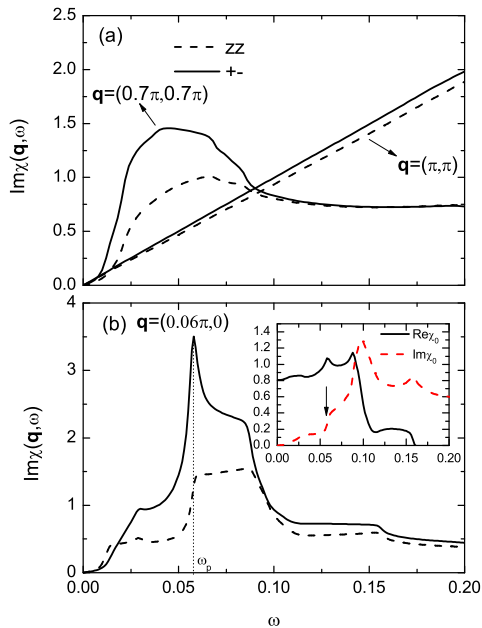}
\caption{(Color online) The imaginary parts of the spin susceptibility as a function of frequency at different momenta for the $p$-wave SC state.
The inset of panel (b) shows the bare spin susceptibility. }
\end{figure}

The spin excitations for the $p$-wave symmetry are depicted in Fig.5. In this case the transverse ($\chi^{+-}$) and longitude $(\chi^{zz})$ spin susceptibility differ with the coherence factors. For both ${\bf q}=(\pi,\pi)$ and ${\bf q}=(0,0)$, one should have $\Delta_{\bf k} \Delta_{{\bf k}+{\bf q}}>0$. Thus generally $\chi^{+-}$ is greater than $\chi^{zz}$. The enhancement of spin response in the SC state can only occur in the $\chi^{+-}$ channel. This can be seen clearly from Fig.5 and the spin susceptibility for $zz$ channel is suppressed by the SC pairing and always smaller than that for $+-$ channel.
Around the $(\pi,\pi)$ momentum, the spin excitation for $\chi^{+-}$ is similar to that of $s$-wave case, i.e., at low frequencies and the momentum $\bf q=(\pi,\pi)$, the spin susceptibility increases as frequency increases. Away from the commensurate momentum, a broad IC peak shows up. At higher frequencies, the commensurate spin excitation is dominant.

Similar to the cases of $d$-wave and $s$-wave, the spin responses at small momenta are incommensurate for all of the frequencies.
The strongest excitation occurs at the momentum $(0.06\pi,0)$ (and its symmetric momenta).
The spin excitation at this momentum is presented in Fig.5(b). As is seen, the spin excitation in $+-$ channel displays a peak structure at the frequency $\omega_p=0.058$. The IC peak intensity is greater than that of normal state while it is not significant enough. Such enhanced spin excitation at this momentum can be explained through the bare spin excitation shown in the inset of Fig.5(b). The origin of the enhancement is similar to that of spin resonance, with a step-like rise for Im$\chi_0$  and a peak structure for Re$\chi_0$ at the frequency $\omega_p$. The difference between the present case with the resonant spin excitation shown in Fig.1(a) is that here Im$\chi_0$ is finite at low frequencies, and thus the pole condition for the complex RPA factor was not satisfied. Therefore, it is a relatively weak feature, while we propose that if it is detected experimentally, it might be used as a signature to support the $p$-wave pairing symmetry.

In summary, we have studied the spin excitations in the newly discovered BiS$_2$ based superconductors based on a two orbital model. Assuming that the SC pairing is mediated by the spin fluctuation, three possible pairing symmetries are suggested. The spin excitations in the SC state is studied in detail. There exist some common features for different pairing symmetries, such as the IC spin excitations at low frequencies.
Meanwhile, distinct spin responses corresponding to different pairing symmetries can be used to probe the pairing symmetry experimentally.

This work was supported by the NSFC under the Grant
No. 11004105, the RGC of Hong Kong under the No.
HKU7055/09P and a CRF of Hong Kong.


\begin{thebibliography}{99}
\bibitem{yos} Yoshikazu Mizuguchi {\it et al.}, arXiv: 1207.3145.
\bibitem{yosh} Yoshikazu Mizuguchi {\it et al.}, arXiv: 1207.3558.
\bibitem{sato} Satoshi Demura {\it et al.}, arXiv: 1207.5248.
 \bibitem{usui} Hidetomo Usui, Katsuhiro Suzuki, and Kazuhiko Kuroki, arXiv: 1207.3888.
 \bibitem{sli} Sheng Li, Huan Yang, Jian Tao, Xiaxin Ding, and Hai-Hu Wen, aeXiv: 1207.4955.
 \bibitem{sing} Shiva Kumar Singh {\it et al.}, arXiv: 1207.5428.
\bibitem{awa} V. P. S. Awana {\it et al.}, arXiv:1207.6845.
\bibitem{kot} Hisashi Kotegawa {\it et al.}, arXiv:1207.6935.
 \bibitem{scal} For a review, see D. J. Scalapino, arXiv:1207.4093.
 \bibitem{jxli} Jian-Xin Li, Phys. Rev. Lett. {\bf 91}, 037002 (2003).

 \bibitem{jxli1} Jian-Xin Li and Z. D. Wang, Phys. Rev. B {\bf 70}, 212512 (2004).
 \bibitem{mai} T. A. Maier and D. J. Scalapino, Phys. Rev. B {\bf 78}, 020514 (2008).
 \bibitem{seo} Kangjun Seo, Chen Fang, B. Andrei Bernevig, and Jiangping Hu, Phys. Rev. B {\bf 79}, 235207 (2009).
 \bibitem{morr} D. K. Morr and D. Pines, Phys. Rev. Lett. {\bf 81}, 1086 (1998).
 \bibitem{zhou} Tao Zhou and Z. D. Wang, Phys. Rev. B {\bf 76}, 094510 (2007).
 \bibitem{gao} Yi Gao, Tao Zhou, C. S. Ting, and Wu-Pei Su, Phys. Rev. B {\bf 82}, 104520 (2010).

 \bibitem{tak} T. Takimoto, T. Hotta, and K. Ueda, Phys. Rev. B {\bf 69}, 104504 (2004).
 \bibitem{kur} K. Kuroki, Y. Tanaka, and R. Arita, Phys. Rev. B {\bf 71}, 024506 (2005).
 \bibitem{zjyao} Z. J. Yao, J. X. Li and Z. D. Wang, New. J. Phys. {\bf 11}, 025009 (2009).


\end{thebibliography}
 \end{document}